\documentstyle[12pt,aaspp,flushrt]{article}     

\begin{document}

\title{ 
L1551NE - Discovery of a Binary Companion
}

\author{G. H. Moriarty-Schieven}
\affil{National Research Council of Canada, Herzberg Institute of
Astrophysics, Joint Astronomy Centre, 660 N. A'ohoku Pl., Hilo, HI
96720 (g.moriarty-schieven@jach.hawaii.edu)} 
 
\author{J. A. Powers}
\affil{University of Hawaii, Hilo, HI 96720  (ja\_powers@yahoo.com)}

\author{H. M. Butner}
\affil{Submillimeter Telescope Observatory, Steward Observatory,
University of Arizona, Tucson, AZ  85721 (hbutner@as.arizona.edu)} 

\author{P. G. Wannier}
\affil{Jet Propulsion Laboratory, MS 169-506, 4800 Oak Grove Dr., Pasadena, 
CA \,\, 91109 (peter.wannier@jpl.nasa.gov)}

\and

\author{Jocelyn Keene}
\affil{California Institute of Technology, MS 320-47, Pasadena, CA
91125  (jkeene@sirtfweb.jpl.nasa.gov)}

\begin{abstract}

L1551NE is a very young (class 0 or I) low-mass protostar located
close to the well-studied L1551 IRS5.  We present here evidence, from
1.3mm continuum interferometric observations at $\sim$1'' resolution,
for a binary companion to L1551NE.  The companion, whose 1.3mm flux
density is $\sim$1/3 that of the primary component, is located 1.43''
($\sim$230 A.U. at 160pc) to the southeast.  The millimeterwave
emission from the primary component may have been just barely
resolved, with deconvolved size $\sim$0.82"x0.70" ($\sim$131x112
A.U.).  The companion emission was unresolved ($<$100 A.U.).   The
pair is embedded within a flattened circum-binary envelope of size
$\sim$5.4'' $\times$ 2.3'' ($\sim 860 \times 370$ A.U.).  The masses
of the three components (i.e. from the cicumstellar material of the
primary star and its companion, and the envelope) are approximately
0.044, 0.014 and 0.023 M$_{\odot}$ respectively. 

\end{abstract}


\pagebreak

\section{Introduction}

Young T Tauri stars in Taurus have been found to have a high incidence
of multiplicity, with the fraction of close ($\lesssim$100A.U.)
companions found to be 0.40 $\pm$ 0.08 (Ghez et al. 1993; Leinert et
al. 1993; Simon et al. 1995; Patience et al. 1998), using speckle and
occultation techniques. Similar surveys of the Hyades cluster found a
smaller fraction of 0.30 $\pm$ 0.06, and a still lower fraction of
0.14 $\pm$ 0.03 for G dwarfs in the solar neighborhood (Patience et
al. 1998), suggesting an evolutionary effect and/or environmental
effects during formation.  Deeply embedded class I and class 0
protostars cannot be surveyed using similar techniques, since they are
not visible at optical wavelengths and are generally too deeply
embedded even at infrared
wavelengths. Millimeter/sub-millimeter-wavelength interferometry at
sub-arcsecond resolution can resolve close companions if they have
circumstellar disks.  For example, L1551 IRS5 was recently shown to be
a binary based on 7mm interferometric observations at the VLA
(Rodr\'iguez et al. 1998). We present here evidence that L1551 NE also
has a binary companion.  

L1551NE (B1950 $4^h28^m50.5^s$ +$18^{\circ}02'10"$ (Draper,
Warren-Smith \& Scarrott 1985)) is a young stellar object in the L1551
molecular cloud, at a distance of 160 pc (Snell 1981).  Discovered by
Emerson et al. (1984) from IRAS data, it is the second brightest
embedded source in the Taurus complex after L1551 IRS5, with L$_{bol}
\sim 6 L_{\odot}$.  It has a molecular outflow (Moriarty-Schieven,
Butner \& Wannier 1995).  The radial density distribution of L1551NE
has been modeled by Barsony \& Chandler (1993) from 800${\mu}m$
images, and by Butner et al. (1995) from 100${\mu}m$ and 200${\mu}m$
observations.  Both found that the density distribution implied by the
radial intensity profile was much shallower than the $n(r) \sim
r^{-1.5}$ predicted by the ``inside-out'' collapse model of Terebey,
Shu \& Cassen (1984).  Moriarty-Schieven, Butner \& Wannier (1995)
have suggested that L1551 NE may be a class 0 source. 

\section{Observations}

The observations were taken with the Owens Valley Radio Observatory 
millimeter-wave array 1994 November 12 and 1997 January 17, at a
wavelength of 1.3mm (230.65 GHz).  The 1994 observations were taken
with the array (consisting at the time of six 10m antennae) in the
compact ``$l$'' configuration (maximum baseline 115m, minimum baseline
30m), and in 1997 the data were taken using the extended ``$h$''
configuration (maximum baseline 241.7m, minimum baseline 35m).  The
effective resolution of the synthesized beam was $\sim$2'' and
$\sim$1'' for the two configurations.   

The observations were made in ``snapshot'' mode, with a series of ten
minute integrations interspersed with observations of other sources
and the phase calibrators (0528+134 and 3C84), to yield a total
integration time of $\sim$1 hour in the ``$l$'' configuration and
$\sim$2 hours in the ``$h$'' configuration.  The system temperatures
were $\sim$1200K and $\sim$1600K respectively.  Neptune was used as
the primary flux calibrator.  The continuum bandwidth is 1GHz. 

The data were reduced using the standard reduction package ``mma'',
and then exported to AIPS which was used to generate and clean images
(using the `IMAGR' task and uniform weighting).  The clean box
included only a small area covering the main source and the apparent
binary companion.  The angular resolution of the cleaned image was
1.29'' $\times$ 1.07'' at position angle -74$^{\circ}$. 

\section{Data}

Figure 1 presents a contour image of L1551NE.  Clearly seen are the
primary source (henceforth source A) at (B1950) 04$^h$28$^m$50.552$^s$
18$^{\circ}$02'09.85'', and another weaker source 1.43'' to the
southeast (source B).  Emission also appears to extend to the north
and to the east of the primary source, and surrounding source B. 

To verify that we are seeing real sources and not phase instabilities,
we generated images using the same techniques of another source, IRAS
04169+2702, which was observed during the same two days and
interspersed with L1551NE.  Phase errors would be manifested as
``anomalous'' sources or structures.  No such anomalous sources are
seen in the image of IRAS 04169+2702.  In addition, we cleaned the
``$l$'' configuration data separately from the ``$h$'' configuration
data.  The high-resolution data clearly had two peaks, while the
low-resolution data showed an extended disk-like structure with long
axis through the line joining the two sources.  Thus we believe that
the structure seen here in L1551NE is real. 

In Table 1 we present the positions, sizes and flux densities of the
sources.  The single-dish flux density at 1.35mm is also shown. 

\section{Discussion}

There are three distinct components apparent in the image shown in
Figure 1; a brighter, possibly extended source at the field center
(source A), a weaker, probably point-like source (B) 1.43'' south-east
of A, and diffuse, low-level emission which appears to surround both
sources and extend $\sim$2'' to the northwest and east of A. 

Sources A and B were fit with elliptical gaussians using the {\em
AIPS} task JMFIT.  The primary source A was found to have a size
1.53''x1.28'' (i.e. it may have been slightly resolved with a size
$\sim 2 \sigma$ larger than the beamsize), peak intensity $\sim$ 0.33
Jy and integrated intensity  $\sim$ 0.47 Jy (i.e. $\sim$40\% 
($\sim 3\sigma$) larger than the peak intensity).  If it is slightly resolved
(deconvolved size $\sim$ 0.82''x0.70''), then its size is 
$\sim$131x112 A.U. at a distance of 160pc.  Source B is located 1.43'' 
(229 A.U. at 160pc) south-east of A, and its size and intensity are 
consistent with it being unresolved, i.e. $<$100 A.U.  We estimate the 
mass within each source using $M_D = \frac{S_\nu 
D^2}{B_\nu(T)}~\frac{4}{3} \frac{a\rho}{Q_\nu}$, and display these in 
Table 1.  We used the integrated flux densities, assumed a distance of 
160 pc (Snell 1981), used dust temperature of 42K (Moriarty-Schieven 
et al. 1994), and    
dust emissivities from Hildebrand (1983) $\frac{4}{3}
\frac{a\rho}{Q_\nu} = 0.1 (\lambda/250)^{\beta}$~g~cm$^{-2}$, assuming
$\beta$ = 1 (Moriarty-Schieven et al. 1994), and have assumed a gas to
dust mass ratio of 100.  The derived masses are approximately 0.044
and 0.014 M$_{\odot}$ for  the emission from source A and B
respectively, and 0.022M$_{\odot}$ for the envelope.  The total mass
for the two sources plus envelope is $\sim$0.078 M$_{\odot}$.

Rodr\'iguez, Anglada \& Raga (1995) imaged the L1551NE region at
$\lambda$3.5c m and found a continuum source located within
1.5$\sigma$ of our source A.  Rodr\'iguez et al attribute the
$\lambda$3.5cm emission to shocks associated with the outflow
(Moriarty-Schieven, Butner \& Wannier 1995).  Thus source A is the
likely origin of the L1551NE outflow.  Rodr\'guez et al. found a
possible second source located $\sim$0.6'' east of A, which is not
coincident with our source B, for which no $\lambda$3.5cm emission was
detected.  If their source 0.6'' east of A is real, this would suggest
that L1551NE is at least a triple star system.  However, another
protostar capable of generating an outflow jet should have had a
circumstellar disk large enough to be detected by our observations.
Possibly the eastern $\lambda$3.5cm source is due to a jet from A but
offset from the source, or is a background object. 

The single-dish flux density (Butner et al. 2000) found using the 14m
JCMT telescope (FWHM 20'')) is not significantly different from the
total intensity found in our image of L1551NE (primary beam FWHM
28'').  Thus only a small amount, if any, of the single-dish flux can
have been resolved out by the interferometer.  However, considerable
extended emission was seen at $\sim$850 $\mu$m by Barsony \& Chandler
(1993) and Moriarty-Schieven et al. (1999), and at 200 $\mu$m by
Butner et al. (1995).  Barsony \& Chandler and Butner et al. modeled
this extended emission as an envelope whose radial density
distribution decreases very slowly with distance from the protostar.

Weak, low-level emission can be seen extending $\sim$1-2'' to the
north and east of source A, and perhaps encompassing source B.  This
extended structure has a disk-like appearance, of dimension
$\sim5''\times2''$ ($\sim$800x300pc with long axis at position angle
$\sim$-12$^{\circ}$.  This is roughly perpendicular to the axis of the
conical reflection nebula emanating from L1551NE (Draper et al. 1985;
Hodapp 1995), and hence of the molecular outflow (Moriarty-Schieven et
al. 1995).  This disk-like structure may represent a circum-binary
disk.

\acknowledgements

JAP was supported by a Hawai'i Space Grant College Fellowship which is
funded by the NASA Undergraduate Space Grant Fellowship program.  The
Owens Valley millimeter-wave array is supported by NSF grant
AST-96-13717.

\newpage

\begin{table}[t]
\begin{center}
\caption{Source Parameters} \label{src_tab}
\begin{tabular}{ll}
\hline \hline
\multicolumn{2}{l}{\bf Source A} \\
Peak Intensity 	 & 0.333 $\pm$ 0.022 Jy/beam \\
Integrated Intensity & 0.473 $\pm$ 0.049 Jy \\
Position (B1950) & 04$^h$28$^m$50.559$^s$ 18$^{\circ}$02'09.91'' ($\pm$ 0.1'') \\
Size             & 1.53'' $\times$ 1.28'' $\pm$ 0.1'' P.A. 104$^{\circ}$ $\pm$ 15$^{\circ}$\\
``Deconvolved'' Size             & 0.82'' $\times$ 0.70'' (131 $\times$ 112 A.U. at 160pc) \\
Mass$^a$	 & 0.044 M$_{\odot}$ \\
\hline
\multicolumn{2}{l}{\bf Source B} \\
Peak Intensity 	 & 0.146 $\pm$ 0.022 Jy/beam \\
Integrated Intensity & 0.197 $\pm$ 0.048 Jy \\
Position (B1950) & 04$^h$28$^m$50.604$^s$ 18$^{\circ}$02'08.64'' ($\pm$ 0.17'') \\
Mass$^a$	 & 0.014 M$_{\odot}$ \\
\hline
\multicolumn{2}{l}{\bf Circumbinary Disk} \\
Integrated Intensity   & 0.233 $\pm$ 0.048 Jy \\
Size		 & 5.51'' ($\pm$ 0.44'') $\times$ 2.59'' ($\pm$
0.22'') P.A. 2$^{\circ}$ ($\pm$ 14$^{\circ}$) 
\\
Deconvolved Size & 5.4'' ($\pm$ 0.3'') $\times$ 2.3'' ($\pm$
0.2'') P.A. 2$^{\circ}$ ($\pm$ 10$^{\circ}$) \\
Mass$^a$	& 0.022 M$_{\odot}$ \\
\hline
\multicolumn{2}{l}{\bf Total Integrated Intensity} \\
Integ. Intensity & 0.851 $\pm$ 0.084 Jy  \\
Mass$^a$        & 0.079 M$_{\odot}$ \\
\hline
\multicolumn{2}{l}{\bf Single Dish Intensity$^b$} \\
19'' FWHM beam   & 0.83 $\pm$ 0.03 Jy \\
Mass$^a$	 & 0.078 M$_{\odot}$ \\
\hline
\multicolumn{2}{l}{$^a$Assuming T$_d$=42K, M$_g$/M$_d$=100, D=160pc.} \\
\multicolumn{2}{l}{$^b$From Butner et al. (2000).  Obtained with 14m
JCMT (20'' FWHM)).} \\
\end{tabular}
\end{center}
\end{table}

\newpage

\begin{figure}[h]
\centering
  \vspace*{15cm}
  \leavevmode
  \includegraphics{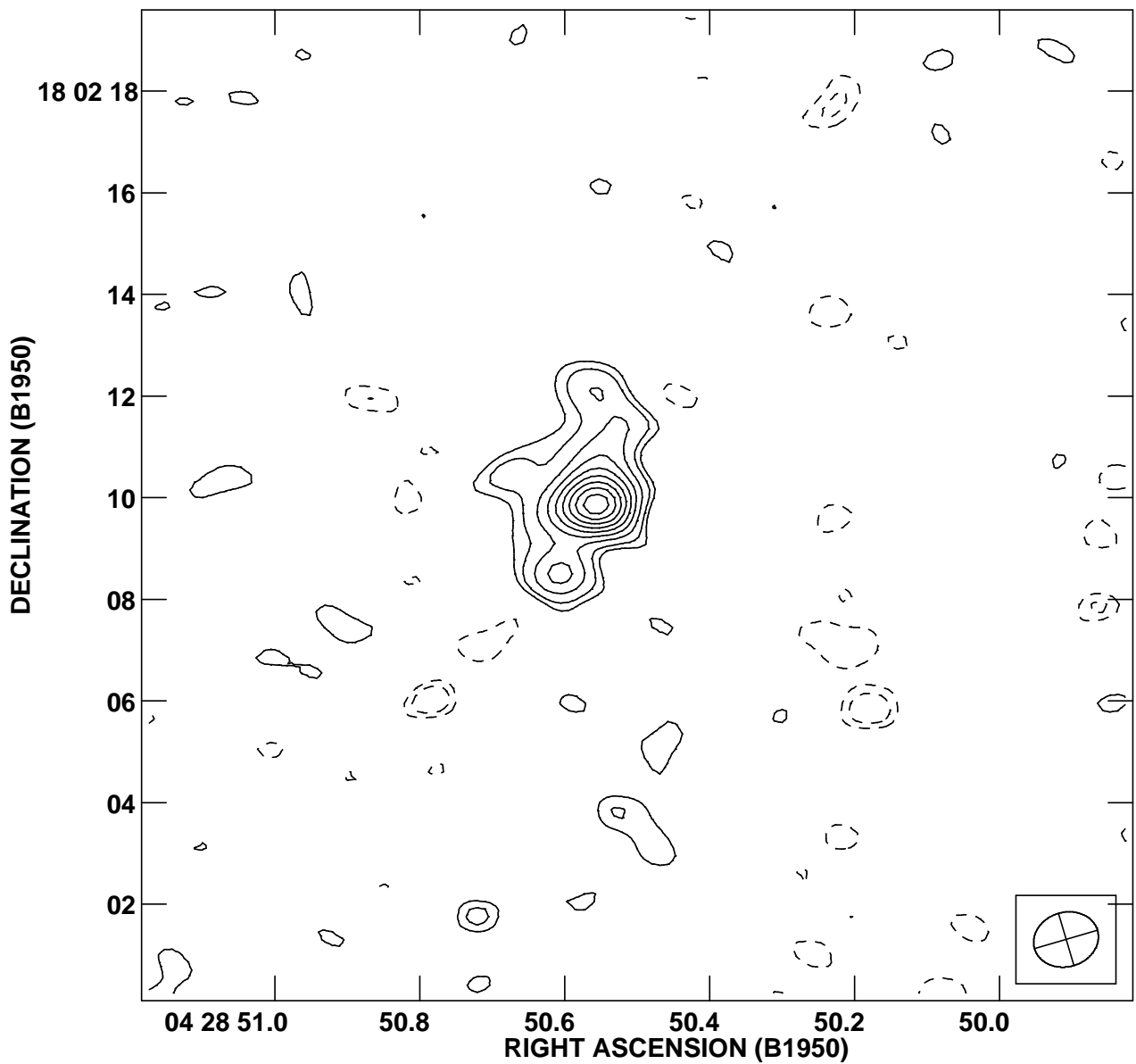}
\caption{Contour image of L1551NE.  Source A is the brighter source at
the field center, B is 1.43'' to the southeast.  The two lowest
positive contours are at 0.049 and 0.07 Jy/beam, and increment at
intervals of 0.035 Jy/beam.  The negative contours are -0.049 an -0.07
Jy/beam.  RMS noise of the image is 0.023 Jy/beam.  The synthesized
beam (FWHM) is shown in a box in the lower-right corner of the image.}

\end{figure}

\end{document}